\documentclass{elsart}
\usepackage{natbib,psfig}
\begin{document}
\runauthor{Hill \& Rawlings}
\begin{frontmatter}
\title{The TOOT Survey: status and early results}
\author[gary]{Gary J. Hill}
\author[steve]{Steve Rawlings}

\address[gary]{McDonald Observatory, University of Texas at Austin, 
	RLM 15.308, TX 78712, USA}
\address[steve]{Astrophysics, Oxford University, Keble Road, Oxford OX1 3RH, UK}
\begin{abstract}
The TexOx-1000 (TOOT) radio source redshift survey is designed to find and
study {\it typical} radio-loud active galaxies to high redshift.
They are typical in the same sense that
L$^*$ galaxies are typical of galaxies in the optical.
Previous surveys
have only included the most luminous, rare objects at and beyond the peak of
activity at z$\sim$2, but in going a factor of 100 fainter than the 3C survey, 
and in assembling a large sample, TOOT probes for the first time the
objects that
dominate the radio luminosity density of the universe at high redshift. 
Here we describe the current
status of the TOOT survey and draw preliminary conclusions about the redshift
distribution of the radio sources.  
So far, $\sim$520 of the 1000 radio sources have
redshifts, with $\sim$440 of those in well-defined, complete, sub-regions of 
the survey. For these we find a median redshift of z=1, but the measured 
redshift distribution has a deficit
of objects with z$\sim$2, when compared to predictions based on 
extrapolating luminosity functions constrained by higher-flux-density samples.
These are the more luminous objects that
usually show emission lines, and which
should not be missed in the survey unless they are heavily reddened.
The deficit may be real, but it would not be too surprising to find a
population of faint, reddened radio sources at z$\sim$2-3
among the TOOT sources yet to have accurate redshifts.
\end{abstract}
\end{frontmatter}

\section{Introduction and Motivation}
The TexOx-1000 (TOOT) radio source (RS) survey is an ambitious collaboration
between Texas and Oxford Universities to assemble the largest sample of
RS, in
order to probe typical radio-loud active galaxies to high redshift.  The TOOT
survey
has made use of UK and Texas facilities to identify and measure the
redshifts of
a large sample drawn from the low-frequency 7C survey 
with S{$_{151 {\rm MHz}}$}$>$0.1 Jy.  
It is already the largest RS redshift survey in existence that extends to 
high redshift and is unbiased by optical selection criteria.

\begin{figure}
\psfig{file=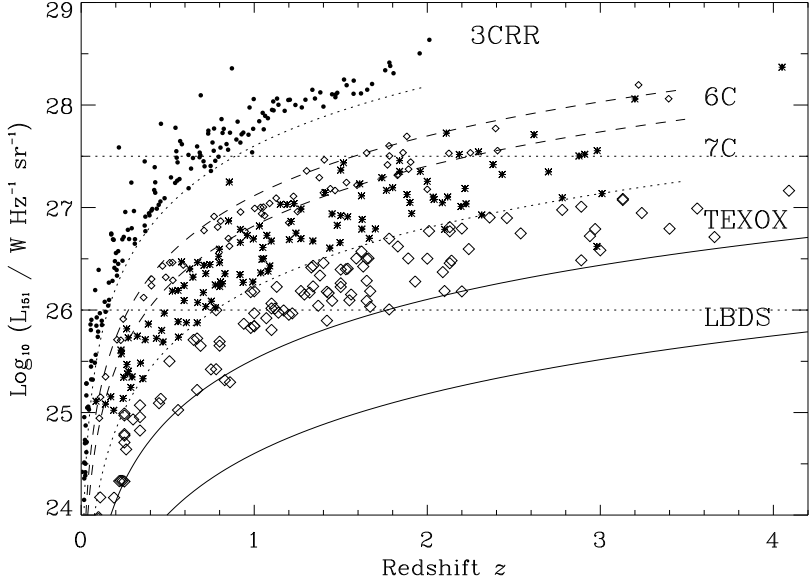,angle=0,width=12cm}
\caption{The Luminosity - Redshift plane for low frequency radio surveys 
including TOOT. 
The TOOT survey is labeled as TEXOX.
The curved lines indicate the flux-density limits of the
surveys, and the horizontal dotted lines show the range of radio luminosities 
that dominate the radio luminosity density of the universe (see text).
The objects indicated for TOOT are a subset of the sample discussed here.}
\label{fig1}
\end{figure}

Fig. 1 shows the radio-luminosity (L$_{151}$) 
versus redshift diagram for various complete
low-frequency RS redshift surveys, including TOOT.  Selection at low frequency
is essential in order to avoid orientation-dependent beaming effects
that distort
the source counts at high frequency.  
The 3CRR \cite{LRL}, 6CE \cite{6CE}, and 7CRS \cite{7CRS} surveys
have a total of 356 RS, but still suffer from small number statistics and the
fact that
even the 7CRS does not reach the radio luminosity function (RLF) break at
z$>$1. As a result, there is still controversy about the form of the RLF
evolution
at high redshift, specifically whether there is a cut-off beyond z=2
\cite{DP,J01,W01}.
The TOOT
survey on the other hand probes low-enough L$_{151}$ at high redshift, and is
sufficiently large, to resolve the form of the radio source population (RSP)
evolution.  Even at high redshift, TOOT contains objects with L$_{151}$
typical of 3CR galaxies at z$\sim$0.3, allowing direct comparison of properties
as a function of redshift without the usual problems of the redshift-luminosity
degeneracy.
The lower flux-density LBDS survey \cite{LBDS} misses the bulk of the
luminosity density at all redshifts, and is dominated by very low L$_{151}$ RS.
The TOOT survey is the realisation of a survey proposed by Willott {\it et al.}
(2001, W01 \cite{W01}), and it has several key aims:

\begin{itemize}
\item Probe the existence of a redshift cutoff for the high-luminosity 
sources at z$>$2. We will be able to finally get away from
small number statistics and measure the magnitude of any
high-$z$ space density decline.
\item Establish whether the low-luminosity population evolves at all.
At present we cannot discriminate reliably between
constant space density and a rise by an order-of-magnitude: 
TOOT will establish the evolution to $\pm 0.1$ dex.
\item  Probe large scale structure. Steep spectrum RS are ideal sparse
tracers of LSS, having high bias and being easy to observe over large volumes
\cite{KB,H01}.
\end{itemize}

TOOT is the first survey to study the sources responsible for the bulk of the
radio-luminosity density of the universe at high redshift
(between the horizontal dotted lines in Fig. 1).  
They are typical in the same sense that
L$^*$ galaxies are typical of galaxies in the optical. 
Higher-L$_{151}$
RS are too rare to be important sources of energy input in the evolution of
galaxies, as the RLF is steep.  Radio source 
jet energy is coupled efficiently to the gas in the ISM and ICM and {\it may}
account for $\sim$10\% of the thermal energy of hot gas in clusters,
a significant fraction of the total \cite{SR}.
The injection of jet energy 
may also influence surrounding galaxies by heating the gas in 
the cluster environment.
Study of the RS responsible for the radio luminosity density of the universe is 
hence of great interest.

At the TOOT flux-density limit,
the RSP divides about equally into low- and high-luminosity sources. 
The low-L$_{151}$ RS (below log(L$_{151}$)=26 WHz$^{-1}$Sr$^{-1}$ in Fig. 1)
do not show emission lines and have both FRI and FRII morphology,
while the high-L$_{151}$ ones have emission lines and FRII morphology. 
In TOOT we trace the low-L$_{151}$ population to z$\sim$2. 
W01 have found a good fit to the RSP evolution for S$_{151}$$>$0.5 Jy
by combining a strongly evolving high-L$_{151}$ population with a moderately
evolving low-L$_{151}$ population. TOOT can directly test the predictions of
such models.

\typeout{SET RUN AUTHOR to \@runauthor}

\section{Properties and Observational Status of TOOT}

\begin{table*}
\begin{center}
\caption{The Six TOOT Regions}
\begin{tabular}{l c c | l c c}
\hline 
                 & RA(J2000)   & dec(J2000)  & & RA(J2000)   & dec(J2000)  \\
\hline 
TOOT00   &  00:30       &   +35   &  TOOT02   &  02:00        &  +40     \\
TOOT08   &  08:25         & +27   &  TOOT12   &  12:50       & +33       \\
TOOT16   &  16:40        &  +45   &  TOOT18   &  18:00       &   +63     \\
\hline 
\end{tabular}
\end{center}
\end{table*}

TOOT targets six regions spread in RA in the northern hemisphere at
declinations accessible to the Hobby-Eberly Telescope (HET),
containing a
total of $\sim$1000 RS with S$_{151}$$>$0.1 Jy (Table 1).
We have high resolution radio maps of all sources, 
either from FIRST \cite{FIRST} or our own VLA observations.
Most of the full sample has an identification, either from the
Digital Sky Survey, R-band imaging from the McDonald Observatory 2.7 m
Smith Reflector, or K-band imaging from UKIRT.  

We are currently about half way to our goal of measuring redshifts for the
complete sample of 1000 RS.  Table 2 summarizes the current completeness.
Spectroscopy has been
obtained with the HET Marcario Low Resolution Spectrograph 
\footnote{The HET is operated by McDonald Observatory on behalf of
The University of Texas at Austin, the Pennsylvania State
University, Stanford University,
Ludwig-Maximilians-Universit\"at M\"unchen, 
and Georg-August-Universit\"at G\"ottingen.
The Marcario LRS is a joint project of the HET
partnership and the Instituto de Astronom\'ia
de la Universidad Nacional Aut\'onoma de M\'exico.} \cite{LRS},
the ISIS spectrograph 
\cite{ISIS}
on the WHT, or the IGI spectrograph 
\cite{IGI} 
on the Smith Reflector.  
These facilities
complement each other well: the ISIS has excellent blue response enabling the
observation of Ly$\alpha$ emission for z$>$1.6 for the higher-luminosity sources
that show emission lines; the HET has the aperture to
measure absorption-line redshifts for the faint red galaxies with lower L$_{151}$.
For the high-redshift FRII population, spectra were often obtained
blind, with the slit aligned to the radio axis.

\begin{table*}
\begin{center}
\caption{Current Status of the TOOT Survey Spectroscopy}
\begin{tabular}{l c c c l}
\hline 
Region &       S$_{151}$ (mJy)    &       area (sq.$^o$) & \# spec &status  \\
\hline 
TOOT00s  & $>$100   &  4    & $\sim$50/54  & $\sim$90\% complete spectroscopy \\
TOOT08   & $>$100   & 20  & $\sim$140/153  & $\sim$90\% complete spectroscopy \\
TOOT12   & $>$100      & 25   & $\sim$260/265  & $\sim$complete spectroscopy \\
TOOT16\_100    & 100-200        & 63    & $\sim$40/121 & $\sim$complete IDs  \\
TOOT18 	& $>$4000       & 570   & 37/37  & complete spectroscopy \\
\hline 
Totals         &       &       & $\sim$440/472       & TOOT00s, 08, 12\\
                       &       &       & $\sim$520/630        & full sample \\
\hline 
\end{tabular}
\end{center}
\end{table*}

As we have progressed, we have endeavoured to obtain completeness
in sub-regions of the survey (TOOT00, 08, 12, and 18, see Table 2),  
and now have spectroscopy for half the sample.
Obviously, a fraction of the redshifts will be unreliable or spurious $-$ it
is an unavoidable feature of such studies.  The significantly smaller
6C and 7CRS surveys have achieved
$>$90\% reliable redshifts.  We estimate that we have reliable redshifts
for 70-80\% of the sample, with most of
the unreliable$/$spurious ones associated with objects in the 0.6$<$z$<$1.6 
redshift range, where most
of the galaxies do not show emission lines, and are faint, resulting in
poor S$/$N ratio.  As the HET
continues to improve in performance, we will increase the 
completeness for this part of the sample,
as large aperture is needed for these observations.  At z$>$1.6, most
objects have sufficiently high L$_{151}$ to show emission lines.
Here the ambiguities center on objects with single-line redshifts, but K-band
photometry ensures that alternative line IDs can usually be ruled out reliably.
We are continuing to obtain K-band imaging to improve redshift 
estimates for the faintest objects.

\section{The Preliminary Redshift Distribution}

\begin{figure*}
\psfig{file=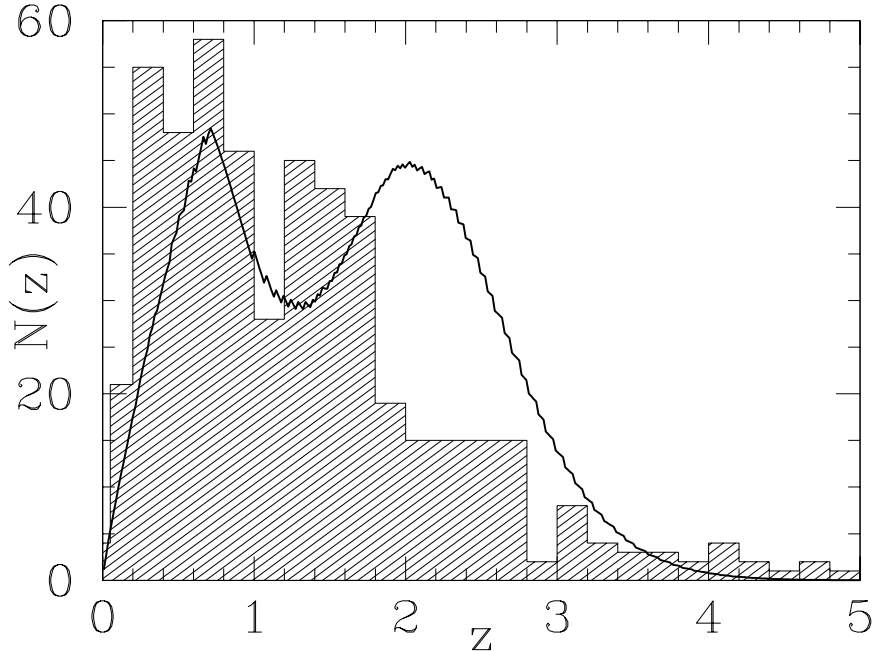,angle=270,width=12cm}
\caption{Preliminary redshift distribution for half the TOOT survey.
The solid curve is the W01 model C prediction, normalised appropriately
(see text).}
\label{fig2}
\end{figure*}

Fig. 2 shows the redshift distribution for 
442
sources in the complete TOOT00s, 08, and 12, fields.
The $\sim$30 sources without spectroscopy are not shown.  The median
redshift is z=1.  The W01 model C,
with an intermediate high-redshift drop-off, computed for 
$\Lambda$=0.7, $\Omega$=0.3 
cosmology is overplotted at appropriate normalisation, for comparison.  
The double-peaked shape of the model prediction is due to the
inflexion in the RLF between the two populations.
The first conclusion from Fig. 2 is that the
TOOT survey has been worthwhile: the model predictions are close to the
observations for z$<$1, but
differ significantly at z$>$2.  
It is difficult
to simply ascribe the disagreement to spurious/incomplete redshifts
as there are too few objects without
redshifts, and we do not expect spurious redshifts for z$>$1.6, 
as long as the objects show emission lines.
However, it is possible that the faint red radio galaxy population, for which
redshifts will be most unreliable, contains
examples of highly-reddened radio galaxies (EROs \cite{ERO}) 
that should be at z$>$2, but are erroneously
ascribed lower redshifts.  In order to fully account for the
discrepancy, the EROs would be 20\%
of the sample.  This is higher than the 8\% found in the 7CRS, for example
\cite{ERO},
and it seems unlikely that we would have just missed those with 2$<$z$<$3.  
Indeed, the 7CRS EROs all lie between 1$<$z$<$2.
On the other hand, our investigation of the properties of TOOT and 7CRS
RS at z$\sim$3 \cite{z3} indicates that they are all reddened quasars, so
a significant fraction of reddened, faint RS might be expected.
With the current state of the spectroscopy we cannot rule out
a significant population of EROs in TOOT at z$\sim$2-3, particularly 
as these objects have similar L$_{151}$ to the lower-redshift 7CRS EROs.
At higher redshift, the agreement with the W01 model C is quite good,
indicating that any high-redshift cutoff is gradual.  
The factor of 10 increase in the space density of the low-L$_{151}$ population
to z$\sim$1 in the W01 models appears to give a good qualitative match to N(z)
for z$<$1.5. 
We will be comparing the results from TOOT to other models \cite{JG}
as we progress.

{\bf Acknowledgements}
We thank Julia Riley for obtaining the radio data on which TOOT is based.
We thank Steve Croft, Kate Brand, Joe Tufts, Marcel Bergmann, and Pamela Gay
for their efforts in obtaining and reducing the TOOT data. 
Chris Willott has had significant influence on the design of the TOOT survey. 
We also thank the staffs of the WHT,
HET, UKIRT, and McDonald observatories, for their
assistance in obtaining the dataset.
This material is based in part upon work supported by the Texas
Advanced Research Program under Grant No. 009658-0710-1999.

\end{document}